\documentclass[11pt,a4paper]{article}
\usepackage{jheppub}
\usepackage[T1]{fontenc}
\usepackage{amssymb}
\usepackage{mathtools}

\usepackage{stackengine}

\usepackage{scalerel}

\usepackage{hyperref}

\usepackage{epsfig}
\usepackage{amsmath,latexsym,amssymb}
\usepackage{graphicx}
\usepackage[latin1]{inputenc}
\usepackage{braket}
\usepackage{pdfsync}

\usepackage{framed}

\usepackage{mathtools}

\def\be{\begin{equation}}
\def\ee{\end{equation}}
\def\ba{\begin{eqnarray}}
\def\ea{\end{eqnarray}}

\newcommand{\la}{\lambda}

\newcommand{\bz}{\bar{z}}

\newcommand{\zbar}{\bar{z}}

\def\wbar{\bar w}

\def\d{\delta}
\def\D{\Delta}
\def\e{\epsilon}

\def\m{\mu}
\def\n{\nu}
\def\om{\omega}

\def\cM{{\cal M}}

\def\cO{{\cal O}}

\newcommand{\comment}[1]{}

\newcommand{\eea}{\end{eqnarray}}

\author{
Angelos Fotopoulos${}^{1,2}$ and
Tomasz R.\ Taylor${}^{1}$ \\[0.5cm]
 $^1${\it Department of Physics \\
  Northeastern University, Boston, MA 02115, USA}\\[0.2cm]
 $^2${\it Department of Natural Sciences\\  Assumption College, Worcester, MA 01609, USA}
 }

\title{\boldmath Primary Fields in Celestial CFT  \unboldmath}

\abstract{The basic ingredient of CCFT holography is to regard four-dimensional amplitudes describing conformal wave packets as two-dimensional conformal correlation functions of the operators associated to external particles. By construction, these operators transform as quasi-primary fields under $SL(2,\mathbb{C})$ conformal symmetry group of the celestial sphere. We derive the OPE of the CCFT energy-momentum tensor with the operators representing gauge bosons and show that they transform as Virasoro primaries under diffeomorphisms of the celestial sphere.}


\bigskip

\begin{document}
\maketitle
\section{Introduction}
Two-dimensional celestial conformal field theory (CCFT) has been recently considered as a candidate for a holographic description of four-dimensional space-time. The developments leading to CCFT \cite{Strominger:2017zoo,Pasterski:2019msg} build upon earlier proposals of flat holography \cite{deBoer:2003vf} and extended BMS symmetries of asymptotically flat spacetime \cite{Barnich:2010eb}. In this framework,
four-dimensional scattering amplitudes are represented as conformal field correlators, with their Lorentz symmetry realized as the  $SL(2,\mathbb{C})$ conformal symmetry group of celestial sphere (${\cal C S}^2$) \cite{Pasterski:2016qvg,Pasterski:2017ylz,Schreiber:2017jsr}. The origin of this connection is a natural identification of the kinematic variables describing asymptotic directions of external particles with the points on ${\cal C S}^2$. The energies are dualized by using Mellin transforms into dimensions of conformal field operators \cite{Pasterski:2017kqt}. The Mellin transforms of Yang-Mills amplitudes are well-defined but the amplitudes involving gravitons usually diverge in the high energy region. Nevertheless,  as demonstrated  in \cite{Stieberger:2018edy}, this problem does not appear in string amplitudes due to their super-soft ultraviolet behaviour.

At this point, it is rather premature to call CCFT a ``theory''. Very few things are known but the emerging picture has some novel and intriguing features.
The CCFT correlators describe the scattering processes of conformal wave packets characterized by $SL(2,\mathbb{C})$ conformal weights $(h,\bar h)$. There is a quasi-primary operator associated to each packet. The conformal spin $h-\bar h$ is identified with helicity, while the condition of normalizability restricts the dimension $\D=h+\bar h$ to the principal continuous series with $\Delta=1+i\lambda$ \cite{Pasterski:2017kqt}. Four-dimensional Lorentz generators are identified with $SL(2,\mathbb{C})$ generators  while four-dimensional translations are realized as dimension-shifting operators \cite{Stieberger:2018onx}.

The correlators representing four-dimensional Einstein-Yang-Mills scattering amplitudes have a complicated structure. This is quite puzzling because, at least at the tree level, Yang-Mills amplitudes exhibit a remarkable simplicity when written in the helicity basis \cite{tt}. There is, however, one limit in which these correlators reveal some interesting features of CCFT.  In the so-called conformal soft limit of $\Delta=1$ \cite{Donnay:2018neh}, the gauge boson operators can be identified with two-dimensional currents representing particles that are soft in the traditional sense (zero energy) \cite{FFT2019}. As a result, the correlators involving such soft insertions are determined by current Ward identities. The conformal soft limit has been subject of many recent studies \cite{FFT2019,Pate:2019mfs,Nandan:2019jas,
Adamo:2019ipt,Puhm:2019zbl,Banerjee:2019aoy,Banerjee:2019tam,
Guevara:2019ypd}.

In a recent work \cite{FFT2019}, we looked at the soft limit from a different perspective. As Ward identities of CCFT, soft theorems reflect the symmetry algebra -- in two dimensions completely determined by the operator product expansion (OPE).
We derived the OPE for the operators representing gauge bosons and used standard CFT techniques to derive Ward identities describing the conformally soft limit.

In CCFT, with the scattering amplitudes transformed into conformal correlators, the limit of coinciding points of ${\cal C S}^2$ corresponds to the limit of coinciding momentum directions, that is to the collinear limit. Hence in principle, it should  be possible to extract not only the products of gauge boson operators, as it was done in Ref.\cite{FFT2019}, but all OPEs from well-known collinear limits. One particularly important set of OPEs involves the energy momentum tensor. It has been proposed \cite{Kapec:2016jld,Cheung:2016iub} that the energy-momentum tensor of CCFT is the $\Delta=2$ operator obtained by taking a shadow transform of the $\Delta=0$ operator associated to the graviton. The analysis of Ref.\cite{Donnay:2018neh} gave further evidence to this proposal. In this work, we put it  on solid foundations by studying the products of this $\Delta=2$ shadow operator with gauge boson operators. We show that the gauge boson operators have the expected OPE with the energy-momentum tensor, appropriate for Virasoro primary fields.
\label{intro}
\section{Preliminaries: conformal wave packets of gravitons}
A general light-like momentum vector of a massless particle can be written as
\be\label{momparam}
p^\mu= \omega q^\mu, \qquad q^\mu={1\over 2} (1+|z|^2, z+\bz, -i(z-\bz), 1-|z|^2)\ ,
\ee
where $\omega$ is the light-cone energy and $q^\mu$ is a null vector -- the direction along which the massless state propagates. It is parametrized by complex $z$ which is identified as a point on ${\cal C S}^2$.
On ${\cal C S}^2$, the $SL(2,\mathbb{C})$ Lorentz group acts as the global conformal symmetry group:
\be\label{SLz}
z\to {a z+b\over c z+d}, \qquad ad-bc=1.
\ee
The usual spin two graviton plane waves are
\be\label{planewave}
\e_{\mu\nu}^{\ell}(p) e^{\mp i |p_0| X^0  \pm i \vec{p} \cdot \vec{X}}, \qquad \ell=\pm 2
\ee
where $\e_{\mu\nu}^{\ell}(p)$  is the polarization  tensor, $\ell$ is the helicity and the $\pm$ sign in the exponential is used to distinguish between incoming and outgoing solutions.\footnote{From this point, we will be considering outgoing (+) solutions only.} It is convenient to choose a gauge in which
\be\label{gravg}
\e_{\mu\nu}^{\ell}(p) =\partial_Jq^\mu\partial_Jq^\nu, \ee
where $J=z$ for $l={+}2$ and $J=\zbar$ for $l={-}2$. In order to transform the scattering amplitudes into conformal correlators, the basis of plane waves will be changed to the basis of wave packets transforming as quasi-primary fields under $SL(2,\mathbb{C})$. The chiral weights $h$ and $\bar h$ are constrained by $h-\bar h=\ell$, but the dimensions $\Delta=h+\bar h$ are {\em a prori\/} arbitrary.

The quasi-primary wave packets satisfying free spin two wave equations, with $h-\bar h=\ell=\pm 2$ and dimension $\Delta$,  were constructed in \cite{Pasterski:2017kqt}. They are given by
\be\label{eq:spin2confprim}
h^{\D,\ell}_{\m \n}= {1\over 2} {[(-q\cdot X) \partial_J q_\m+ (\partial_J q \cdot X) q_\m][(-q\cdot X) \partial_J q_\n+ (\partial_J q \cdot X) q_\n] \over (-q \cdot X)^{\D+2}}
\ee
They are related to Mellin transforms of plane waves in the following way. First, we apply $\partial^2_J$ to the following identity \cite{Pasterski:2017kqt}:
\be\label{eq:diffeo1}
{q_\m q_\n \over (-q\cdot X)^\D}= {1\over 2} {1\over \D-1} \bigg[ \partial_\mu \Big({q_\nu \over (-q\cdot X)^{\D-1}}\Big) + \partial_\n{q_\m \over (-q\cdot X)^{\D-1}}\bigg]\ .
\ee
As a result,
\be\label{eq:diffeo2}
{(\partial_J q\cdot X)^2 q_\m q_\n \over (-q \cdot X)^{\D+2}}= {-2 \over \D (\D+1)}  {\partial_J q_\m  \partial_J q_\n \over (- q\cdot X)^\D} - {2\over \D+1}  {\partial_J q_\m (\partial_J q\cdot X) q_\n +\partial_J q_\n (\partial_J q\cdot X) q_\m \over (- q\cdot X)^{\D+1}}+ \makebox{diff}
\ee
where diff is a ``pure gauge'' part representing a diffeomorphism.
Next, we use
\be\label{eq:diffeo3}
\partial_J {(\partial_J q_{\{\m}) q_{\n\}} \over (-q\cdot X)^\D}= {1\over 2( \D-1)}  \partial_J\bigg[ \partial_\m {(\partial_J q_\n) \over (-q\cdot X)^{\D-1}} + \partial_\n{(\partial_J q_\m) \over (-q\cdot X)^{\D-1}}\bigg]
\ee
to obtain
\be\label{eq:diffeo4}
{\partial_J q_\m (\partial_J q\cdot X) q_\n+ \partial_J q_\n (\partial q_J\cdot X) q_\m \over (-q\cdot X)^{\D+1}}= -{2\over \D} {\partial_J q_\m \partial_J q_\n \over (-q \cdot X)^\D} + \makebox{diff}
\ee
After combining (\ref{eq:diffeo2}) and (\ref{eq:diffeo4}), we find that the conformal wave packet can be written as
\be\label{eq:spin2confprimnew}
h^{\D,\ell}_{\m \n }=  {1\over 2}{\D(\D-1)\over \D+1} {\partial_J q_\m \partial_J q_\n \over (-q \cdot X)^\D}  + \makebox{diff}
\ee

Note that in two special cases, of $\D=0$ and $\D=1$,  the wave functions (\ref{eq:spin2confprimnew}) represent diffeomorphisms. Pasterski and Shao \cite{Pasterski:2017kqt} showed that for a generic $\Delta$, the wave functions (\ref{eq:spin2confprimnew}) are normalizable only if $\Delta=1+i\lambda$, i.e.\ when the fields belong to the principal continuous series. The amplitudes involving gravitons of the principal series have been discussed in \cite{Stieberger:2018edy}  and their conformally soft limit of $\D=1$ in \cite{Adamo:2019ipt,Puhm:2019zbl}. In the next section, we focus on $\Delta=0$, which is another case of pure diffeomorphism.

The connection to Mellin-transformed plane waves is established in the following way. The Mellin transform of the plane wave (\ref{planewave}) is
\be\label{confprimMellin}
G^{\D,\ell}_{\mu \n} (X^\mu, z, \bz)\equiv \partial_J q_\mu \partial_J q_\nu \int _0^\infty d\omega  \ \omega^{\D-1} e^{\mp i \omega q \cdot  X -\e \omega\ .}
\ee
The conformal wave function (\ref{eq:spin2confprimnew}) can be written as\footnote{We skip some irrelevant $i^\D$ factors.}
\be\label{confprimexpr3}
h^{\D,\ell}_{\m \n }=
f(\D)G^{\D,\ell}_{\mu \n} +\makebox{diff}
\ee
where
\be\label{gdef}f(\D)=
{1\over 2}{\D(\D-1) \over \Gamma(\D+2)} \ee

In the next section, we will discuss the shadow transforms of amplitudes involving $\Delta=0$ gravitons,\footnote{Actually, $\D=0$ modes are pure diffeomorphisms, so it would be more appropriate to call them ``Goldstone'' modes instead of gravitons.}
which should correspond to the insertions of the energy-momentum tensor.
 Note that the respective Mellin transforms are accompanied by the normalization factors
 $f(\Delta)$ that vanish in the $\Delta=0$ limit, therefore in order to obtain finite results, the amplitudes must be singular. Indeed, finite OPE coefficients will be extrated from such  singular limits.

\section{OPE of the energy-momentum tensor with primary fields}
We wish to discuss two statements. The first statement is that, as anticipated in \cite{Kapec:2016jld,Cheung:2016iub}, the CCFT energy-momentum tensor $T(z)$
is given by a shadow transform \cite{Osborn:2012vt} of the operator ${\cal G}_{\D=0}(z,\zbar)$ associated to the $\D=0$ limit $(h=-1,\bar h=1)$ of the quasi-primary graviton wave packet (\ref{eq:spin2confprim}) with helicity $\ell={-}2$, hence associated to a non-trivial effect of a pure diffeomorphism, see Eq.(\ref{eq:spin2confprimnew}). More precisely,
\be\label{eq:shadow}
T(z)={3!\over 2\pi} \int d^2 z'{1\over (z'-z)^{4}} \, {\cal G}_{\D=0}(z',\bz')
\ee
The second statement is that the gauge boson operators $\cO(z,\zbar)$ associated to quasi-primary spin one wave functions transform  as conformal primaries under ${\cal C S}^2$ diffeomorphisms.
 This is equivalent to the  OPE
 \be \label{opeto} T(z)\cO(w,\wbar)={h\over (z-w)^2}\cO(w,\wbar)+{1\over z-w}\partial_w\cO(w,\wbar)+\dots \ ,
 \ee
where $h$ is the chiral weight of $\cO $. In this section, we prove the second statement, assuming that the first one is true. To that end, we will be considering tree-level Einstein-Yang-Mills amplitudes involving one graviton and an arbitrary number $N$ of gauge bosons. In Ref.\cite{Stieberger:2016lng} it was shown that such amplitudes can be expressed as linear combinations of pure Yang-Mills amplitudes involving $N{+}1$ gauge bosons, but our discussion will not rely on this relation.

We will proceed in the following steps. We will start from the Mellin transformation converting the standard amplitude into a  correlator of the graviton operator ${\cal G}_\D(z',\zbar')$ with $N$ gauge boson operators $\cO$ inserted at arbitrary points $z_n$. Next, we will convert ${\cal G}_\D$ into its shadow $\widetilde{\cal G}_\D(z,\zbar)$ and extract collinear singularities in the limit of $z$ coinciding with one of $z_n$. After taking the $\D=0$ limit, we will extract the OPE coefficients.

The CCFT correlator of  $N$ gauge bosons and one helicity $\ell=-2$ graviton is given by
\ba\label{eq:graviton-2amp}
\Big\langle \prod_{n=1}^N\cO_{\ell_n}(z_n,\zbar_n){\cal G}_\D(z',\zbar')\Big\rangle&=&\Big(  \prod_{n=1}^N  \int d \omega_n  \ \omega_n^{\D_n-1} \Big) \ f(\D)\int d\om'\, \om'^{\D-1} \d^{(4)}\big(\sum_{n=1}^N  \om_n q_n+\om' q'\big)\nonumber\\[1mm] \label{eq:graviton-2amp}
 &&~~~~~\times
 \cM_{\ell_1\dots \ell_N, \ell=-2}(\omega_n, z_n, \bz_n; \om', z',\bz')
\ea
where $\cM_{\ell_1\dots \ell_N, \ell=-2}$ is the Feynman's matrix element for the scattering of $N$ gauge bosons with helicities $\ell_n$ and one helicity $\ell=-2$ graviton. We omit gauge indices and gauge operator normalization factors because they are not relevant to our discussion, and stick to one partial amplitude corresponding say to the canonical ordering inside the Chan-Paton trace factor. Recall that the dimensions of gauge boson operators are fixed to $\D_n=1+i\la_n$.

Next, we take the shadow transform. Since we are interested in the $\D=0$ limit and $f(\D)$ is already of order $\D$,  we can evaluate this transform as if $\D$  were vanishing:\footnote{The shadow formula valid for arbitrary $(h,\bar h)$ can be found in Ref.\cite{Osborn:2012vt}.}
\be\label{eq:shadow-2amp}
\Big\langle \prod_{n=1}^N\cO_{\ell_n}(z_n,\zbar_n)\widetilde{\cal G}_\D(z,\zbar)\Big\rangle
= {3!\over 2\pi}\int d^2 z' {1\over (z-z')^4} \Big\langle \prod_{n=1}^N\cO_{\ell_n}(z_n,\zbar_n){\cal G}_\D(z',\zbar')\Big\rangle
\ee
Since our goal is to extract the OPE, we are interested in the pole behavior of the above expression in the limit when $z$ approaches one of other points, say $z \to z_N$. A short reflection reveals that poles will appear in the integration region where $z'\to z_N$. Looking back at Eq.(\ref{eq:graviton-2amp}), we see that this is the collinear limit of 
$\cM_{\ell_1\dots \ell_N, \ell=-2}(p_n,p')$
 when the graviton's four-momentum $p'=\omega'q'$ becomes parallel to the momentum $p_N=\omega_Nq_N$ of the $N$th gauge boson. The matrix element $\cM$ is singular in this limit in a rather peculiar way. The pole terms $(\bz'-\bz_N)^{-1}$ are always accompanied by factors $(z'-z_N)$ so they are singular only if $z$ and $\bar z$ are considered as independent variables, otherwise it is just a phase ambiguity reflecting an azimuthal asymmetry \cite{Bern:1998xc}. This singularity is important for the shadow transform (\ref{eq:shadow-2amp}). It can be extracted \cite{btf} from the Feynman diagrams in which the graviton is radiated away by the $N$th gauge boson. One finds
 \ba \cM(1,2,\dots,N^-\!,\,p'^{--})\label{ampcol1}&=& {z'-z_N\over \bz'-\bz_N} { \om_{P} \over \om'} \cM(1,2, \dots, P^-)+\dots\ ,\\ \cM(1,2,\dots,N^+\!,\,p'^{--})\label{ampcol2}&=& {z'-z_N\over \bz'-\bz_N} { \om_N^2 \over \om'\om_P} \cM(1,2, \dots, P^+)+\dots\ , \ea
 where we omitted non-singular terms. We use the notation of Ref.\cite{tt}, with $\pm$ superscripts referring to particle helicities. Here, $P$ is the combined momentum of the collinear pair:
 \be \label{psum}P=p_N+p'=\omega_Nq_N+\om'q'=\omega_Pq_P\ ,\ee
 with
 \be\label{psum1}\om_P=\om_N+\om'\, \qquad q_P=q_N=q'~~(z_P=z_N=z'~,\bz_P=\bz_N=\bz')\ .\ee

 We begin with the case of $\ell_N=-1$, with the collinear limit given in (\ref{ampcol1}). After inserting it into
Eqs.(\ref{eq:graviton-2amp},\ref{eq:shadow-2amp}), we obtain
\ba
\Big\langle \prod_{n=1}^{N-1}&&\!\!\!\!\!\cO_{\ell_n}(z_n,\zbar_n)
\cO_{-1}(z_N,\zbar_N)\widetilde{\cal G}_\D(z,\zbar)\Big\rangle=
\Big(  \prod_{n=1}^{N-1}  \int d \omega_n \omega_n^{\D_n-1} \Big)\times \nonumber\\&&
 {f(\D)\over 2\pi}\int d\om_N\, \om_N^{\D_N-1}\int d\om'\, \om'^{\D-2}\, \om_P\int d^2 z' {3!\over (z-z')^4}\,{z'_N\over \bz'_N}\,\times\nonumber\\
 &&~~~~\d^{(4)}\big(\sum_{n=1}^{N-1}  \om_n q_n+\om_Pq_P\big) \cM(1,2, \dots, P^-)\ , \label{inte1}
 \ea
 where $z'_N\equiv z'-z_N$ and $\bz'_N\equiv\bz'-\bz_N$.
We integrate by parts over $z'$ two times:
\be\label{inte2}\int d^2 z' {3!\over (z-z')^4}{z'_N\over \bz'_N}\longrightarrow \int d^2 z' {1\over (z-z')^2}
 \bigg[ \partial^2_{z'}\Big({z'_N\over \bz'_N} \Big)+ 2 \partial_{z'} \Big({z'_N\over \bz'_N} \Big)\partial_{z'}+\Big( {z'_N\over \bz'_N}\Big)\partial^2_{z'} \bigg].
\ee
The first term yields
\be\label{inte3}
\partial^2_{z'}\Big({z'_N\over \bz'_N} \Big)=2\pi\delta^{(2)}(z'-z_N)\ \ee
and the integral becomes localized at $z'=z$, giving rise to a double pole of the correlator (\ref{inte1}) at $z=z_N$. In the second term, with
\be\label{inte4}\partial_{z'}\Big({z'_N\over \bz'_N}\Big)={1\over \bz'_N}\ee
we integrate by parts one more time:
\be\label{inte5}\int {d^2 z' \over (z-z')^2}{2\over \bz'_N}\partial_{z'}\longrightarrow -\int d^2 z' \bigg[{4\pi\over z-z'}\delta^{(2)}(z'-z_N)
\partial_{z'}+{2\over \bz'_N(z-z')}\partial_{z'}^2\bigg] .\ee
Here, the first term gives rise to a single pole, while the second term, although it is logarithmically divergent at $z=z_N$, does not contribute any poles. Actually, the third term in (\ref{inte2}) contains a similar logarithmic divergence. These two logarithmic terms disappear, however, in the $\Delta= 0$ limit because they contain $\partial_{z'}^2$ acting on the remaining part of the integrand. As explained below, each $\partial_{z'}$ derivative brings one power of $\omega'$,  therefore the corresponding $\omega'$ integrals are finite. In the $\Delta=0$ limit, they are suppressed by {$f(\Delta)$} normalization factor (\ref{gdef}) which is of order ${\cal O}(\Delta)$.

In Eq.(\ref{inte5}), the derivative $\partial_{z'}$ on the right hand side acts on the remaining part of the integrand, that is the momentum-conserving delta function and the matrix element $\cM$, c.f.\ Eq.(\ref{inte1}). Both these functions depend on $p'=\omega'q'(z')$ through the combination $P=p'+p_N$, see Eqs.(\ref{psum},\ref{psum1}). For any such function
\be\label{eq:colspin3}
\partial_{z'}= {\om'\over \om_P} \partial_{z_P}
\ee
when, as in our case, the derivative is evaluated at $q'=q_P$.\footnote{Actually, $\cM$ depends on the momentum spinor $\lambda_P$. Nevertheless, since this amplitude is extracted by factorizing on the gauge boson pole in the $(p_N,p')$ channel, c.f.\ Eqs.(\ref{ampcol1},\ref{ampcol2}), the dependence on $\lambda_P$ is reduced to a simple polarization-dependent factor, while the other factor depends on vector $P$. After checking these factors, one finds that Eq.(\ref {eq:colspin3}) remains valid.} After putting this altogether, we find that after the shadow transform, Eq.(\ref{inte1}) yields
\ba
\Big\langle \prod_{n=1}^{N-1}&&\!\!\!\!\!\cO_{\ell_n}(z_n,\zbar_n)
\cO_{-1}(z_N,\zbar_N)\widetilde{\cal G}_\D(z,\zbar)\Big\rangle=
\Big(  \prod_{n=1}^{N-1}  \int d \omega_n \omega_n^{\D_n-1} \Big)\times ~~\label{intf1} \\
 f(\D)\bigg[&&\!\!\!\!{1\over (z-z_N)^2}\int d\om_N\, \om_N^{\D_N-1}\int d\om'\, \om'^{\D-2}\, \om_P\,\d^{(4)}\big(\sum_{n=1}^{N-1}  \om_n q_n+\om_Pq_P\big) \cM(1,2, \dots, P^-)\nonumber\\ -&&\!\!\!\!{2\over z-z_N}\int d\om_N\, \om_N^{\D_N-1}\int d\om'\, \om'^{\D-1}\, \partial_{z_P}\Big\{\d^{(4)}\big(\sum_{n=1}^{N-1}  \om_n q_n+\om_Pq_P\big) \cM(1,2, \dots, P^-)\Big\}\bigg]
 \nonumber
 \ea
 Now we can change the energy integration variables from $(\omega_N,\om')$ to $(\omega_P=\omega_N{+}\om',\om')$:
\be\label{changei}
\int_0^\infty d \omega_N\int_0^\infty d \omega'\cdots ~\longrightarrow\int_0^\infty d \omega_P\int_0^{\omega_P}d \omega'\cdots
\ee
and integrate over graviton's energy $\omega'$, with the result
\ba
\Big\langle \prod_{n=1}^{N-1}\cO_{\ell_n}(z_n,\zbar_n)&&\!\!\!\!
\cO_{-1}(z_N,\zbar_N)\widetilde{\cal G}_\D(z,\zbar)\Big\rangle=
\Big(  \prod_{n=1}^{N-1}  \int d \omega_n \omega_n^{\D_n-1} \Big)\int d\om_P\om_P^{\D_N+\D-1}\times \nonumber\\
 f(\D)\bigg[&&\!\!\!\!{B(\D_N,\D-1)\over (z-z_N)^2}\d^{(4)}\big(\sum_{n=1}^{N-1}  \om_n q_n+\om_Pq_P\big) \cM(1,2, \dots, P^-)~~\label{intf2} \\ \,\,-&&\!\!\!2\,{B(\D_N,\D)\over z-z_N}\, \partial_{z_P}\Big\{\d^{(4)}\big(\sum_{n=1}^{N-1}  \om_n q_n+\om_Pq_P\big) \cM(1,2, \dots, P^-)\Big\}\bigg],
 \nonumber
 \ea
where $B$ denotes Euler's beta function. At this point, in order to turn the shadow operator $\widetilde{{\cal G}}_\D(z,\bz)$ into the energy-momentum tensor $T(z)$, we take the $\D=0$ limit.
With the factor
$f(\D)$ given in Eq.(\ref{gdef}),
\be f(\D)B(\D_N,\D-1)\to {\D_N-1\over 2}\ ,\qquad f(\D)B(\D_N,\D)\to -{1\over 2}\label{flimit}\ .\ee
Note that on the r.h.s.\ of Eq.(\ref{intf2}), the energy of the $N$th gauge boson is $\omega_P$ and its momentum is $P=\omega_Pq_P$, with $q_P=q_N$. Furthermore, the amplitude and its derivative are evaluated at $z_P=z_N$. As a result,
\ba
\Big\langle \prod_{n=1}^{N-1}\cO_{\ell_n}(z_n,\zbar_n)&&\!\!\!\!
\cO_{-1}(z_N,\zbar_N)T(z)\Big\rangle=
\Big(  \prod_{n=1}^{N}  \int d \omega_n \omega_n^{\D_n-1} \Big)\times \nonumber\\
 \bigg[&&\!\!\!\!{h_N\over (z-z_N)^2}\d^{(4)}\big(\sum_{n=1}^{N}  \om_n q_n\big) \cM(1,2, \dots, N^-)~~\label{intf3} \\ \,\,+&&\!\!\!{1\over z-z_N} \partial_{z_N}\Big\{\d^{(4)}\big(\sum_{n=1}^{N}  \om_n q_n\big) \cM(1,2, \dots, N^-)\Big\}\bigg],
 \nonumber
 \ea
 where we used the fact that for $\ell_N=-1$, $h_N=(\D_N-1)/2$.

The OPE (\ref{opeto}) follows from the $z\to z_N$ singularities of
the energy-momentum tensor correlator (\ref{intf3}). A similar proof can be carried out in the case of $\ell_N=+1~ [h_N=(\D_N+1)/2]$, by using the collinear limit (\ref{ampcol2}).
\section{Conclusions}
The change of basis from plane waves to conformal wave packets, accomplished by using Mellin transformations, converts Yang-Mills amplitudes into celestial amplitudes. The basic idea of CCFT holography is to regard these amplitudes as the correlation functions of the operators associated to external particles. By construction, these operators must transform as quasi-primaries under the (global) $SL(2,\mathbb{C})$ inherited from four-dimensional Lorentz symmetry. It is by no means obvious that these operators transform as Virasoro primaries because ${\cal C S}^2$ diffeomorphism are generated by {\em a priori\/} unknown CCFT energy-momentum tensor $T(z)$. In this work, guided by Refs.\cite{Kapec:2016jld,Cheung:2016iub}, we assumed that $T(z)$ is given by the shadow of the $\Delta=0$ graviton operator representing pure (large) diffeomorphisms. We showed that under this assumption, the operators representing gauge bosons do indeed transform as Virasoro primaries.

As mentioned before, the assumed form of $T(z)$ was mainly based on the arguments involving soft theorems, in the framework of quantum field theory formulated in curvilinear coordinates in which ${\cal C S}^2$ diffeomorphisms appear as a part of  larger BMS symmetry. It would be very enlightening to discuss this symmetry in a purely CCFT setup, by studying the OPE of $TT$ products. This OPE has already been discussed in the context of double-soft limits in Ref.\cite{Distler:2018rwu}. In our approach, it should emerge from the collinear limit of two shadow operators. Unfortunately, unlike in the case of a single shadow, understanding the collinear behaviour of
Einstein-Yang-Mills amplitudes is not sufficient for extracting the collinear singularities of shadow products. We hope to address this issue in the future.
\vskip 1mm
\leftline{\noindent{\bf Acknowledgments}}
\vskip 1mm
\noindent
We are grateful to Wei Fan for collaborating at an early stage of this work and for checking some computations. We are also grateful to Ignatios Antoniadis for asking all the right questions. T.R.T.\ thanks Zygmunt Lalak and the Institute of Theoretical Physics of the University of Warsaw for their kind hospitality.
This material is based in part upon work supported by the National Science Foundation
under Grant Number PHY--1913328.
Any opinions, findings, and conclusions or recommendations
expressed in this material are those of the authors and do not necessarily
reflect the views of the National Science Foundation.


\begin{thebibliography}{99}

\bibitem{Strominger:2017zoo}
  A.~Strominger,
  {\it Lectures on the Infrared Structure of Gravity and Gauge Theory},
Princeton University Press (2018)
  [arXiv:1703.05448 [hep-th]].
\bibitem{Pasterski:2019msg}
  S.~Pasterski,
  ``Implications of Superrotations,''
  arXiv:1905.10052 [hep-th].


\bibitem{deBoer:2003vf}
  J.~de Boer and S.~N.~Solodukhin,
  ``A Holographic reduction of Minkowski space-time,''
  Nucl.\ Phys.\ B {\bf 665} (2003) 545 (2003)
  [hep-th/0303006].
\bibitem{Barnich:2010eb}
  G.~Barnich and C.~Troessaert,
  ``Aspects of the BMS/CFT correspondence,''
  JHEP {\bf 1005} (2010) 062
  [arXiv:1001.1541 [hep-th]].



\bibitem{Pasterski:2016qvg}
  S.~Pasterski, S.~H.~Shao and A.~Strominger,
  ``Flat Space Amplitudes and Conformal Symmetry of the Celestial Sphere,''
  Phys.\ Rev.\ D {\bf 96} (2017) no. 6, 065026
  [arXiv:1701.00049 [hep-th]].

  \bibitem{Pasterski:2017ylz}
  S.~Pasterski, S.~H.~Shao and A.~Strominger,
  ``Gluon Amplitudes as 2d Conformal Correlators,''
  Phys.\ Rev.\ D {\bf 96} (2017) no. 8, 085006  [arXiv:1706.03917 [hep-th]].

\bibitem{Schreiber:2017jsr}
  A.~Schreiber, A.~Volovich and M.~Zlotnikov,
  ``Tree-level gluon amplitudes on the celestial sphere,''
  Phys.\ Lett.\ B {\bf 781} (2018) 349
  [arXiv:1711.08435 [hep-th]].


\bibitem{Pasterski:2017kqt}
  S.~Pasterski and S.~H.~Shao,
  ``Conformal basis for flat space amplitudes,''
  Phys.\ Rev.\ D {\bf 96} (2017) no. 6, 065022
  [arXiv:1705.01027 [hep-th]].



\bibitem{Stieberger:2018edy}
  S.~Stieberger and T.~R.~Taylor,
  ``Strings on Celestial Sphere,''
  Nucl.\ Phys.\ B {\bf 935} (2018) 388
  [arXiv:1806.05688 [hep-th]].

\bibitem{Stieberger:2018onx}
 S.~Stieberger and T.~R.~Taylor,
  ``Symmetries of Celestial Amplitudes,''
  Phys.\ Lett.\ B {\bf 793}, 141 (2019)
  [arXiv:1812.01080 [hep-th]].
\bibitem{tt}  T.R.~Taylor,
  ``A Course in Amplitudes,''
Phys.\ Rept.\  {\bf 691}, 1 (2017).
[arXiv:1703.05670 [hep-th]].

\bibitem{Donnay:2018neh}
  L.~Donnay, A.~Puhm and A.~Strominger,
  ``Conformally Soft Photons and Gravitons,''
  JHEP {\bf 1901} (2019) 184
  [arXiv:1810.05219 [hep-th]].


\bibitem{FFT2019}
  W.~Fan, A.~Fotopoulos and T.~R.~Taylor,
  ``Soft Limits of Yang-Mills Amplitudes and Conformal Correlators,''
  JHEP {\bf 1905} (2019) 121
  [arXiv:1903.01676 [hep-th]].


\bibitem{Pate:2019mfs}
  M.~Pate, A.~M.~Raclariu and A.~Strominger,
  ``Conformally Soft Theorem in Gauge Theory,''
  arXiv:1904.10831 [hep-th].

\bibitem{Nandan:2019jas}
  D.~Nandan, A.~Schreiber, A.~Volovich and M.~Zlotnikov,
  ``Celestial Amplitudes: Conformal Partial Waves and Soft Limits,''
  arXiv:1904.10940 [hep-th].

\bibitem{Adamo:2019ipt}
  T.~Adamo, L.~Mason and A.~Sharma,
  ``Celestial amplitudes and conformal soft theorems,''
  arXiv:1905.09224 [hep-th].

\bibitem{Puhm:2019zbl}
  A.~Puhm,
  ``Conformally Soft Theorem in Gravity,''
  arXiv:1905.09799 [hep-th].


\bibitem{Banerjee:2019aoy}
  S.~Banerjee, P.~Pandey and P.~Paul,
  ``Conformal properties of soft-operators - 1 : Use of null-states,''
  arXiv:1902.02309 [hep-th].

\bibitem{Banerjee:2019tam}
  S.~Banerjee and P.~Pandey,
  ``Conformal properties of soft operators - 2 : Use of null-states,''
  arXiv:1906.01650 [hep-th].
\bibitem{Guevara:2019ypd}
  A.~Guevara,
  ``Notes on Conformal Soft Theorems and Recursion Relations in Gravity,''
  arXiv:1906.07810 [hep-th].

\bibitem{Kapec:2016jld}
  D.~Kapec, P.~Mitra, A.~M.~Raclariu and A.~Strominger,
  ``2D Stress Tensor for 4D Gravity,''
  Phys.\ Rev.\ Lett.\  {\bf 119} (2017) no.12,  121601
  [arXiv:1609.00282 [hep-th]].

\bibitem{Cheung:2016iub}
  C.~Cheung, A.~de la Fuente and R.~Sundrum,
  ``4D scattering amplitudes and asymptotic symmetries from 2D CFT,''
  JHEP {\bf 1701} (2017) 112
  [arXiv:1609.00732 [hep-th]].




\bibitem{Osborn:2012vt}
  H.~Osborn,
  ``Conformal Blocks for Arbitrary Spins in Two Dimensions,''
  Phys.\ Lett.\ B {\bf 718} (2012) 169
  [arXiv:1205.1941 [hep-th]].

\bibitem{Stieberger:2016lng}
  S.~Stieberger and T.~R.~Taylor,
  ``New relations for Einstein-Yang-Mills amplitudes,''
  Nucl.\ Phys.\ B {\bf 913}, 151 (2016)
  [arXiv:1606.09616 [hep-th]].

\bibitem{Bern:1998xc}
  Z.~Bern, L.~J.~Dixon, M.~Perelstein and J.~S.~Rozowsky,
  ``One loop n point helicity amplitudes in (selfdual) gravity,''
  Phys.\ Lett.\ B {\bf 444}, 273 (1998)
  [hep-th/9809160].
\bibitem{btf}
  A. Fotopoulos, S. Stieberger, T.R.\ Taylor and Bin Zhu, ``BMS Algebra from Soft and Collinear Limits,''
 to appear.
\bibitem{Distler:2018rwu}
  J.~Distler, R.~Flauger and B.~Horn,
  ``Double-soft graviton amplitudes and the extended BMS charge algebra,''
  arXiv:1808.09965 [hep-th].

 \end{thebibliography}
\end{document}